\documentstyle[prl,aps,twocolumn,epsf,floats]{revtex}

\newcommand{\di}{\mbox{d}}

\begin{document}

\draft

\twocolumn[\hsize\textwidth\columnwidth\hsize\csname %
@twocolumnfalse\endcsname


\date{\today}

\title{ Spin-fermion model near the quantum critical point: 
one-loop renormalization group results}
 
\author{Ar. Abanov and Andrey V. Chubukov}

\address
{Department of Physics, University of Wisconsin, Madison, WI 53706}


\maketitle

\begin{abstract}
We consider spin and electronic 
 properties of itinerant electron systems, described by 
the spin-fermion model, near 
the antiferromagnetic critical point.
We expand in the inverse number of hot spots in the Brillouin zone, $N$ 
and present the results beyond previously studied $N = \infty $ limit.
We found two new effects: 
(i) Fermi surface becomes nested at hot spots, 
and (ii) vertex corrections give rise to
 anomalous spin dynamics and change the dynamical critical exponent from $z=2$ to $z>2$.
 To first order in $1/N$ we found $z = 2N/(N-2)$ which for a physical $N=8$ 
yields $z\approx 2.67$.
\end{abstract}

\pacs{PACS numbers: 74.20.Fg, 75.20Hr}
]

\narrowtext

The problem of fermions interacting with critical antiferromagnetic spin fluctuations attracts a lot of attention at the moment due to its relevance to both
high temperature superconductors and heavy-fermion materials~\cite{review}.
 The key interest
of the current studies is to understand the system 
behavior near the quantum critical point (QCP) where the magnetic 
correlation length diverges at $T=0$~\cite{scs}.
Although in  reality the QCP is almost always
 masked by either superconductivity or 
precursor effects to superconductivity, 
the vicinity of the QCP can be reached by  
 varying external parameter such as pressure in
 heavy fermion compounds, or doping concentration in cuprates.
 
In this paper, we study the properties of the QCP without taking pairing fluctuations into account. We assume that the singularities
associated with the closeness to the QCP extent up to energies which 
 exceed typical energies associated with the pairing.
This assumption is consistent with the recent calculations of the pairing instability temperature in cuprates~\cite{acf}. 
From this perspective, the understanding
 of the properties of the QCP without pairing correlations 
is a necessary preliminary step for subsequent studies of the pairing problem.

\begin{figure} [t]
\begin{center}
\leavevmode
\epsfxsize=3.0in 
\epsfysize=1.1in 
\epsffile{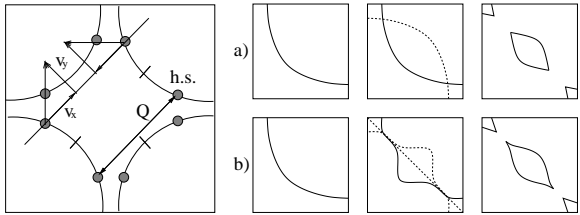}
\end{center}
\caption{
The Fermi surface with hot spots and the directions of Fermi velocities
at hot spots separated by $Q$, and  the evolution of the
 Fermi surface evolution for 
(a) mean-field ($N=\infty)$ SDW theory,
 and (b) finite $N$. In both cases, the doubling of the unit cell due to antiferromagnetic SDW ordering introduces shadow Fermi surface and yields a gap opening near hot spots. At finite $N$, however, the Fermi surface at the quantum critical point 
becomes nested at hot spots due to vanishing of renormalized $v_y$.}
\label{fig1}      
\end{figure}

A detailed study of the antiferromagnetic QCP was performed by 
Hertz~\cite{hertz} and later by Millis~\cite{millis} who chiefly 
focused on finite $T$ properties near the QCP.
They both argued that if the Fermi surface contains hot spots (points 
separated by antiferromagnetic momentum $Q$, see Fig.~\ref{fig1}),
then spin excitations possess 
purely relaxational dynamics with $z=2$. They further argued that in
 $d=2$, $d+z=4$, i.e., the critical theory is at marginal dimension,
 in which case one should expect that spin-spin interaction yields
at maximum  logarithmical corrections to the relaxational dynamics.         
Millis argued~\cite{millis} that this is true provided that 
the effective Ginsburg-Landau functional for spins (obtained by integrating out the fermions) is an analytic function of the spin ordering field.
This is a'priori unclear as the expansion coefficients
in the Ginsburg-Landau functional are made out of particle-hole bubbles and
generally are sensitive to the closeness to quantum criticality due to 
 feedback effect from near critical spin fluctuations 
on the electronic subsystem. Millis however
demonstrated that the quartic term in the 
Ginsburg-Landau functional is governed by high energy fermions and is 
free from singularities. 

In this communication, we, however, argue that the regular Ginsburg-Landau 
expansion is not possible in 2D by the reasons different from those displayed in ~\cite{hertz,millis}. Specifically, we argue that the damping term in the
spin propagator (assumed to be linear in $\omega$ in ~\cite{hertz,millis}) is
by itself made out of a particle hole bubble, and, contrary to $\phi^4$ coefficient, is governed by low-energy fermions. We demonstrate that due to singular vertex corrections, the  
frequency dependence of the spin damping term at the QCP is actually
 $\omega^{1-\alpha}$. In the one loop approximation, we find 
$\alpha \approx 0.25$.  

Another issue which we  study is the form of the renormalized 
quasiparticle Fermi surface near the 
magnetic instability. In a mean-field SDW theory, the Fermi surface 
in a paramagnetic phase is not affected by the closeness to the QCP. 
Below the instability, the doubling of the unit cell induces a 
shadow Fermi surface at $k_F +Q$,
 with the residue proportional to the deviation from criticality. This
gives rise to the opening of the SDW gap near hot spots and eventually (for
a perfect antiferromagnetic long range order) yields a Fermi surface in the 
form of small pockets around $(\pi/2,\pi/2)$ and symmetry related 
points (see Fig.~\ref{fig1}a).
Several groups argued~\cite{ks} 
that this mean-field scenario is modified by fluctuations, 
and the Fermi surface evolution towards hole pockets begins
already within the paramagnetic phase.  
We show that the Fermi surface near hot spots does
evolve as $\xi \rightarrow \infty$, but due to strong fermionic damping (not 
considered in~\cite{ks}), this evolution is a minor effect
which at $\xi = \infty$ only gives rise to a nesting at the hot spots (see Fig. ~\ref{fig1}b). 

The point of departure for our analysis is the spin-fermion model 
which describes low-energy fermions  
interacting with their own collective spin degrees of freedom. 
The model is described by 
\begin{eqnarray}
{\cal H} &=&
  \sum_{{\bf k},\alpha} {\bf v}_F ({\bf k}-{\bf k}_F)  
 c^{\dagger}_{{\bf k},\alpha} c_{{\bf k},\alpha}
+ \sum_q \chi_0^{-1} ({\bf q}) {\bf S}_{\bf q} {\bf S}_{-{\bf q}} +\nonumber \\
&&g \sum_{{\bf q,k},\alpha,\beta}~
c^{\dagger}_{{\bf k+ q}, \alpha}\,
{\bf \sigma}_{\alpha,\beta}\, c_{{\bf k},\beta} \cdot {\bf S}_{\bf -q}\, .
\label{intham}
\end{eqnarray}
Here  $c^{\dagger}_{{\bf k}, \alpha} $ is the fermionic creation operator
for an electron with momentum ${\bf k}$ and spin projection $\alpha$,
$\sigma_i$ are the Pauli matrices,  and 
 $g$ measures the strength of the 
interaction between fermions and their collective bosonic spin
degrees of freedom. The latter are described  by 
 ${\bf S}_{\bf q}$ and are characterized by a bare  spin susceptibility
which is obtained  by  integrating out  high-energy fermions.

This spin-fermion  model can be viewed as 
the appropriate low-energy theory for
Hubbard-type lattice fermion models provided that  spin fluctuations are
the only low-energy degrees of freedom. This model explains a number 
of measured features of cuprates both in the normal and the 
superconducting states~\cite{ac}. Its application to heavy-fermion 
materials is more problematic as in these compounds conduction 
electrons and spins are independent degrees of freedom, and the 
dynamics of spin fluctuations may 
be dominated by local Kondo physics rather than the 
interaction with fermions~\cite{piers}.  

The form of the bare susceptibility $\chi_0 (q)$ is an input 
for the low-energy theory. We 
assume that $\chi_0 (q)$ is  non-singular and peaked at ${\bf Q}$, i.e.,  
$\chi_0 ({\bf q}) = \chi_0/(\xi ^{-2} + ({\bf q}-{\bf Q})^2)$, 
where $\xi$ is the magnetic 
correlation length. 
In principle, $\chi_0$ can 
also contain a nonuniversal frequency dependent term in the form  
$(\omega/W)^2$ 
where $W$ is of order of fermionic bandwidth. We, however, will see that 
for a Fermi surface with hot spots which we consider here, 
this term will be overshadowed by a universal $\omega^{1-\alpha}$ 
term produced by low-energy fermions.

The earlier studies of the spin-fermion model have demonstrated that the
perturbative expansion for both fermionic and bosonic self-energies 
holds in power of $\lambda = 3g^2 \chi_0/(4\pi v_F \xi^{-1})$ where 
$v_F$ is the Fermi velocity at a hot spot. 
This perturbation theory
 obviously does not converge when $\xi \rightarrow \infty$. 
As an alternative to a conventional perturbation theory, we suggested the
expansion in inverse number of  hot spots in the Brillouin zone $N$ 
($=8$ in actual case)~\cite{acf,ac}.
Physically, large $N$ implies that a spin fluctuation has many channels 
to decay into a particle-hole pair,  
which gives rise to a strong ($\sim N$) spin damping rate. 
At the same time, a fermion near a hot spot can only scatter into a single 
hot spot separated by ${\bf Q}$. Power counting arguments than show that a 
large damping rate appears 
in the denominators of the fermionic self-energy and vertex corrections 
and makes them small to the extent of $1/N$. 
The only exception from this rule is the fermionic self-energy due to a 
single spin fluctuation exchange, which contains
 a frequency dependent piece
 without $1/N$ prefactor due to an infrared singularity 
which has to be properly regularized~\cite{chubukov}. 

The set of coupled equations
for fermionic and bosonic self-energies  at $N=\infty$ has 
been solved in ~\cite{chubukov}, and we merely quote the result.  
Near hot spots, we have
\begin{eqnarray}
G_{k}^{-1}(\omega)&=&\omega -\epsilon_k
+\Sigma (\omega ), 
\nonumber \\
\chi (q,\Omega _{m})&=&\chi _{0}\xi
^{2}/(1+({\bf q}-{\bf Q})^{2}\xi ^{2} -i \Pi _\Omega). 
\label{def}
\end{eqnarray}
Here $\epsilon_k = v_x {\tilde k}_x + v_y {\tilde k}_y$, 
where ${\tilde k} = k - k_{hs}$, and $v_x$, $v_y$, which we set to be positive, are the components of the 
Fermi velocity at a hot spot ($v^2_F = v^2_x + v^2_y$). 
The fermionic self-energy $\Sigma_k (\omega)$ and the spin 
polarization operator $\Pi_{\Omega}$ are given by 
\begin{equation}
\Sigma (\omega)=2~\lambda~\frac{\omega}{1+%
\sqrt{1 -\frac{i|\omega|}{\omega _{sf}}}};~\Pi _\Omega=\frac{%
|\Omega|}{\omega _{sf}}  \label{input}
\end{equation}
and $\omega _{sf}=(4\pi/N)~v_x v_y/(g^2 \chi_0 \xi ^{2})$. 

We see from Eq.(\ref{input}) that for $\omega \leq \omega _{sf}$, $%
G(k_{hs},\omega )=Z/(\omega +i\omega |\omega |/(4\omega _{sf}))$, i.e., as
long as $\xi $ is finite, the system preserves the Fermi-liquid behavior at
the lowest frequencies. The quasiparticle residue $Z$ however depends on 
the interaction strength, $Z=(1+ \lambda)^{-1}$,  and 
progressively goes down when the spin-fermion coupling increases. At
larger frequencies $\omega \geq \omega _{sf}$, the system crosses over to a
region, which is in the basin of attraction of the quantum critical point, $%
\xi =\infty $. In this region, $G^{-1}(k_{F},\omega ) \approx 
3 g~(v_x v_y \chi_0 /\pi Nv^2_F)^{1/2}~(i|\omega|)^{1/2}
{\rm sgn}(\omega )$~\cite{chubukov,Mi}.
At the same time, spin propagator has a simple $z=2$ relaxational dynamics unperturbed by strong frequency dependence of the fermionic self-energy~\cite{kadanof}.

\begin{figure} [t]
\begin{center}
\leavevmode
\epsfxsize=2.8in 
\epsfysize=1.1in 
\epsffile{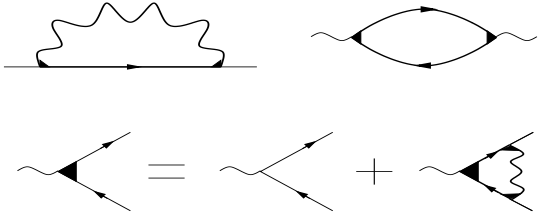}
\end{center}
\caption{ 
The one-loop RG diagrams for the fermionic self-energy and vertex renormalization.
Solid lines are full fermionic propagators, wavy lines are full spin susceptibilities, and black triangles are full vertices. The lowest order diagrams are
obtained by replacing full internal lines and vertices by their $N=\infty$ forms}
\label{fig2}      
\end{figure}

Our present goal is to go beyond $N=\infty$ limit 
and analyze the role of $1/N$ corrections. 
The $1/N$ terms give rise to two new features:
 vertex corrections which renormalize both fermionic and bosonic self-energies, and static 
fermionic self-energy $\Sigma_k$. 
The corresponding diagrams are presented in
 Fig~\ref{fig2}.
 The lowest-order $1/N$ corrections have been calculated 
before~\cite{chubukov,ch-morr}. Both vertex correction and the static
 self-energy are logarithmical in $\xi$:
\begin{eqnarray}
\frac{\Delta g}{g}  &=&  \frac{Q(v)}{N}~ \log {\xi}, 
\label{vert}\\
\Delta \epsilon_k &=& - \epsilon_{k+Q}~
\frac{12}{\pi N}~ \frac{v_x v_y}{v^2_F}~ \log{\xi}
\label{se}
\end{eqnarray}
where $\epsilon_{k+Q} = -v_x {\tilde k}_x + v_y {\tilde k}_y$, and
$Q(v) =(4/\pi) \arctan (v_{x}/v_{y})$
interpolates between 
$Q=1$ for  $v_x = v_y$, and $Q=2$ for $v_y \rightarrow 0$. 

Besides, the $1/N$ corrections also contribute 
$(1/N) \omega \log \xi$ to $G^{-1}_k (\omega)$, but 
this term is negligible compared to $\Sigma (\omega)$ and we neglect it.


We see from (\ref{vert},\ref{se}) that the $1/N$ corrections 
to the vertex and to the velocity of the excitations
are almost decoupled from each other: the velocity renormalization does 
not depend on the coupling strength at all, 
while the renormalization of the vertex depends on the ratio of velocities 
only through a non-singular $Q(v)$.
This is a direct consequence of the fact that the dynamical part of the 
spin propagator is 
 obtained self-consistently within the model. Indeed, the
overall factors in $\Delta\epsilon_k$ and 
$\Delta g/g$ are $g^2 (\omega_{sf} \xi^2)$ where 
$\omega_{sf} \xi^2$ comes from the dynamical part of the spin susceptibility.
Since the fermionic damping is produced by
 the same spin-fermion interaction as the fermionic self-energy, 
$\omega_{sf}$ scales as $1/g^2$, 
and the coupling constant disappears from the r.h.s. of (\ref{vert},\ref{se}). 

The logarithmical dependence on $\xi$ implies that $1/N$ expansion 
breaks down near the QCP, and one has to sum up the series of the 
logarithmical corrections. We will do this in a standard one-loop 
approximation by summing up the series in $(1/N) \log \xi$ but 
neglecting regular 
$1/N$ corrections to each term in the series. We verified that in 
this approximation,  the cancellation of the coupling constant holds 
even when $g$ is a running, scale dependent 
 coupling. This in turn implies that one can separate the velocity 
renormalization 
from the renormalization of the vertex to all orders in $1/N$.

Separating the corrections to $v_x$ and $v_y$ and 
performing standard RG manipulations, we  obtain a set of two RG equations
for the running $v^R_x$ and $v^R_y$
\begin{eqnarray}
\frac{\di v^R_{x}}{\di L}&=&\frac{12}{\pi N} 
\frac{ (v^R_x)^2  v^R_y}{ (v^R_{x})^{2}+ (v^R_{y})^{2}}    \nonumber \\
\frac{\di  v^R_{y}}{\di L}&=&-\frac{12}{\pi N} 
\frac{ (v^R_{y})^2  v^R_x}{ (v^R_{x})^{2}+ (v^R_{y})^{2}}\label{Renorm-velocity}
\end{eqnarray}
where $L = \log \xi$. 
The solution of these equations is straightforward, and yields
\begin{equation}
v^R_x = v_x Z; v^R_y = v_y Z^{-1}; 
Z=\left(1 + \frac{24 L}{\pi N} \frac{v_y}{v_x}\right)^{1/2}
\label{sol}
\end{equation}
where, we remind, $v_x$ and $v_y$ are the bare values of the 
velocities (the ones which appear in the Hamiltonian). 

We see that $v^R_y$ vanishes logarithmically at
$\xi \rightarrow \infty$. This implies that right at the QCP, the
renormalized velocities  at $k_{hs}$ and $k_{hs}+Q$ are antiparallel 
to each other, i.e. the Fermi surface becomes nested at hot spots 
(see Fig~\ref{fig1}b).
This nesting 
%
creates a ``bottle neck effect'' immediately below the criticality as the original and the 
shadow Fermi surfaces approach hot spots with equal derivatives (see Fig. ~\ref{fig1}b). This obviously helps developing a SDW gap at $k_{hs}$ below the 
magnetic instability. However, above the transition, no SDW precursors 
appear at $T=0$.

Another feature of the RG equations (\ref{Renorm-velocity}) is
 that they leave the product $v_x v_y$ unchanged. This is a combination in which velocities appear in $\omega_{sf}$. The fact that $v_x v_y$ is not renormalized  implies that, without vertex renormalization, $\omega_{sf} \xi^2$ remains finite at $\xi = \infty$, i.e., spin fluctuations preserve a simple $z=2$ relaxational dynamics. 

We now consider vertex renormalization. Using again the fact that $g^2 \omega_{sf}$ does not depend on the running 
coupling constant,
 one can straightforwardly extent the second-order result for the vertex renormalization, Eqn (\ref{vert}), 
to the one-loop RG equation
\begin{equation}
\frac{\di g^R}{\di L} =  \frac{Q(v)}{N} g^R 
\label{rgg}
\end{equation}
where $g^R$ is a running coupling constant, and 
$Q(v)$ is the same as in (\ref{vert}) but contain renormalized velocities 
$v^R_x$ and $v^R_y$. 
At the QCP, the dependence on $\xi$ obviously transforms into the dependence
on frequency ($L = \log \xi \rightarrow 
(1/2) \log |\omega_0/\omega|$, where $\omega_0$ is the upper cutoff). 
Using the fact that for $\xi \rightarrow  \infty$,
 $v^R_y/v^R_x \approx N\pi/24 L$  and expanding  $Q(v)$ 
near $v^R_y =0$, we find
 $Q(v) \approx 2 (1 - (2/\pi) v^R_y/v^R_x) = 2 - N/3 L$.
Substituting this result into 
(\ref{rgg}) and solving the differential equation we obtain (${\bar \omega} = \omega/\omega_0$)
\begin{equation}
g^R = g~ |\bar{\omega}|^{-1/N}~|\log {\bar\omega}|^{-1/6}   
\label{gr}
\end{equation}
We see that at the QCP, running coupling constant diverges as $\omega \rightarrow 0$ roughly as $|\omega|^{-1/N}$. Substituting this result into 
the spin polarization operator and using the fact that 
$\omega_{sf} \propto (g^R)^{-2}$ we find that at the QCP, 
\begin{equation}
\Pi_\Omega \propto  |\omega|^{\frac{N-2}{N}}~ |\log \omega|^{-\frac{1}{3}}
\label{chan}
\end{equation}
This result implies that vertex corrections change the
 dynamical exponent $z$ from its mean-field value $z=2$
 to $z = 2N/(N-2)$.
For $N=8$, this yields $z \approx 2.67$ and $\chi (Q,\omega) \propto 
|\omega|^{1-\alpha}$ where $\alpha =0.75$. 

Singular vertex corrections also renormalize the fermionic self-energy
as $\Sigma (\omega) \propto g^R \sqrt{|\omega|}/v_F$. 
Using the results for $g^R$ and $v_F \approx v_x$
we obtain at criticality
\begin{equation}
\Sigma(\omega) \propto |\omega|^{\frac{N-2}{2N}}~  
|\log \omega|^{-\frac{2}{3}}
\label{sian}
\end{equation}
Eqs. (\ref{sol}), (\ref{chan}) and (\ref{sian}) are the central results of the paper.
We see that the singular corrections to the Fermi velocity cause nesting but
do not affect the spin dynamics.
 The corrections to the vertex on the other hand do not affect velocities, but change the dynamical critical exponent for spin fluctuations. 


We now  briefly discuss the form of the susceptibility at finite $T$.
Previous studies have demonstrated~\cite{scs,millis} that the 
scattering of a given spin fluctuation by classical, thermal spin 
fluctuations yields, up to logarithmical prefactors,
$\xi^{-2} \propto u T$, where 
$u$ is the coefficient 
 in the $\phi^4$ term in the Ginsburg-Landau potential. This implies that 
at the QCP, $\chi (Q,\omega) \propto T -i |\omega|$.

We, however, argue that the linear in $T$ and the 
linear in $\omega$ terms have completely different origin: the linear in
$\omega$ term comes from low-energies and is universal, while the linear 
in $T$ term comes from high energies and is model dependent.
This can be understood by 
analyzing the particle-hole bubble at finite $T$. 
We found that as long as one restricts with the linear expansion 
near the Fermi surface,  $\Pi_\Omega$  preserves exactly the same form 
as at $T=0$, to all orders in the perturbation theory. 
 The temperature dependence of $\Pi$ appears only due to a
nonzero curvature of the electronic dispersion and is obviously sensitive
 to the details of the dispersion at energies comparable to the bandwidth. Similarly,  the derivation of the Landau-Ginsburg potential from 
(\ref{intham}) shows~\cite{millis} that 
$u$ vanishes for linearized $\epsilon_k$, 
and is finite only due to a nonzero curvature of the fermionic dispersion. 

The different origins of $T$ and $\omega$ dependences in $\chi 
(Q,\omega)$ imply that the anomalous $\omega^{1-\alpha}$ frequency 
dependence of 
$\chi (Q,\Omega)$ is not accompanied by the 
anomalous temperature dependence of $\chi (Q,0)$ simply because
for high energy fermions, vertex corrections are non-singular.  
This result 
implies, in particular, that our theory does not explain 
anomalous spin dynamics observed in heavy fermion 
\cite{l} despite the similarity in the
exponent for the frequency dependence of $\Pi_\Omega$, because the 
experimental data imply the existence of the $\Omega/T$ scaling in 
$CeCu_{6-x}Au_x$~. More likely, the explanation should involve the 
local Kondo physics~\cite{piers}. 

Finally, we consider how anomalous vertex corrections affect the 
superconducting problem. We and Finkel'stein argued recently~\cite{acf} 
that at $\xi = \infty$, the 
 kernel $K(\omega, \Omega)$ of the 
Eliashberg-type gap equation for the $d-$wave anomalous vertex 
$F(\Omega) = (\pi T/2) \sum_{\omega} K(\omega, \Omega) F(\omega)$ 
behaves as $K (\omega,\Omega)\propto g^2/(v^2_F \Sigma^2 (\omega) 
\Pi_{\Omega -\omega})^{1/2}$
At $N=\infty$, this yields (including prefactor)
$K (\omega,\Omega)= |\omega (\Omega -\omega)|^{-1/2}$. Although
this kernel is qualitatively different from the one in the BCS theory 
because it depends on both frequencies, it still scales as 
inverse frequency due to an interplay between a non-Fermi liquid 
form of the fermionic self-energy and
the absence of the gap in the spin susceptibility which mediates pairing.
We demonstrated in ~\cite{acf} that 
this inverse frequency dependence 
gives rise to a finite pairing instability temperature 
even when $\xi = \infty$.

To check how the kernel is affected by vertex corrections, we substitute
the results for $g^R$, $v_F$, $\Sigma (\omega)$ and $\Pi_\Omega$ 
into $K(\omega,\Omega)$. We find after simple manipulations that 
{\it despite
singular vertex corrections, the kernel in the gap equation still scales 
inversely proportional to frequency}. A simple extension of the 
analysis in ~\cite{acf} then shows that the system still possesses a 
pairing instability at $\xi = \infty$ at a temperature which differs 
from that without vertex 
renormalization only by $1/N$ corrections. 

To summarize, in this paper we considered the properties of the 
antiferromagnetic quantum critical point for itinerant electrons 
by expanding in the inverse number of hot spots in the Brillouin zone $N=8$. 
We went beyond a self-consistent $N=\infty$ theory and found two new effects:
(i) Fermi surface becomes nested at hot spots which is a weak
 SDW precursor effect, and (ii) vertex corrections account for anomalous 
spin dynamics and change the dynamical critical exponent from 
$z=2$ to $z>2$. To first order in $1/N$ we found 
$z = 2N/(N-2)\approx 2.67$. We argued that anomalous
 frequency dependence is not accompanied by anomalous $T$ dependence.

It is our pleasure to thank G. Blumberg, P. Coleman, 
M. Grilli,  A. Finkel'stein, D. Khveshchenko, 
A. Millis, H. von L\"{o}hneysen, J. Schmalian, Q. Si, and A. Tsvelik 
 for useful conversations. 
The research was supported by NSF DMR-9979749.

\end{document}